\documentstyle[11pt,a4]{article}

\newcommand{\vs}[1]{\rule[- #1 mm]{0mm}{#1 mm}}

\newcommand{\be}{\begin{equation}}
\newcommand{\ee}{\end{equation}}

\newcommand{\bea}{\begin{eqnarray}}
\newcommand{\eea}{\end{eqnarray}}

\newcommand{\lp}{\left(}
\newcommand{\rp}{\right)}
\newcommand{\scr}{\scriptstyle}

\newcommand{\om}{\omega}
\newcommand{\PP}{\phi^{\dagger}\phi}
\newcommand{\dV}{\frac{d}{dV}}
\newcommand{\parV}{\frac{\partial}{\partial V}}
\newcommand{\cI}{\oint_{\cal C}\frac{d\om}{4\pi i}}
\newcommand{\Vp}{V^{\prime}}

\newcommand{\pho}{\phi^{(0)}}

\newcommand{\sect}[1]{\setcounter{equation}{0}\section{#1}}

\begin{document}

\begin{titlepage}

\leftline{\Large {ITP-UH-11-95}}
\rightline{\Large {March 1995}}
\leftline{\Large {hep-th/9503185}}

\vs{20}

\begin{center}

{\LARGE {\bf Loop equations for multi-cut \\[.5cm]
             matrix models}}\\[2cm]

{\large G. Akemann}
\footnote{supported by the `Studienstiftung des Deutschen Volkes'} \\
{\em Institut f\"ur Theoretische Physik, Universit\"at Hannover\\
Appelstra{\ss}e 2, 30167 Hannover, Germany}\\
{akemann@itp.uni-hannover.de}\\[.5cm]

\end{center}

\vs{20}

\centerline{ {\bf Abstract}}

The loop equation for the complex one-matrix model with a multi-cut
structure is derived and solved in the planar limit. An iterative
scheme for higher genus contributions to the free energy and the
multi-loop correlators is presented for the two-cut model, where
explicit results are given up to and including genus two. The
double-scaling limit is analyzed and the relation to the one-cut solution
of the hermitian and complex one-matrix model is discussed.

\end{titlepage}

\renewcommand{\thefootnote}{\arabic{footnote}}
\setcounter{footnote}{0}

\sect{Introduction}

\indent

Within the topic of matrix models multi-cut solutions are of considerable
interest. They are intimately related to the existence of multi-critical
points. In order to reach, say, an $m$th critical point, $m-1$ coupling
constants have to be introduced in the matrix potential and adjusted in the
right way. This immediately leads to the possibility of multi-cut
solutions because there can be as many cuts as minima of the
potential.

Unfortunately there is not much known about higher-order contributions
in the topological $1/N$ expansion for multi-cut solutions. The saddle-point
approximation provides only the planar solution, whereas the full
non-perturbative treatment with orthogonal polynomials has been successful
only for one or at most two cuts except for special cases like degenerate
minima of the potential \cite{DE90}. The reason is that the appropriate
ansatz for solving the string equation is not known in general. The
assumption for the recursion coefficients yielding several continuous
functions in the large-$N$ limit does not match with the semi-classical
analysis for higher-order
potentials \cite{LE91}. Numerical studies \cite{SASU91,LE92,JU91,SENE92,
BDJT93} have shown the existence of instabilities in the solution of the
string equation for a variety of different potentials. This phenomenon has
been subsumed under the catchword of `chaos in matrix models'.
There have been attempts to explain the origin of these
oscillations \cite{JU91} but a full understanding is still lacking.

Within the framework of the third method of solving matrix models, the
technique of loop equations \cite{MIG83}, there has been significant
progress during
the last years. Ambj{\o}rn $et$ $al$. \cite{ACM92,AKM92,AMB93} have proposed
a very effective scheme to calculate higher-genus contributions in the
perturbative expansion. Making use of a redefinition from coupling
constants to moments it allows one to determine all multi-loop correlators
order by order in the genus expansion.

The aim of this paper is to demonstrate how this method can be applied to
multi-cut solutions, where the complex matrix model \cite{MOR91} has been
considered for simplicity. The loop equation
\footnote{The approach of \cite{AKM92} adopted here considerably differs
from \cite{CDM92}.}
and the starting point, the planar solution of the one-loop correlator, can
be obtained for any number of cuts. However, for more than two cuts
technical difficulties enter the game via
a new type of equation determining the edges of the cuts.
So the complete iterative solution of the two-cut complex one-matrix model
presented here may be seen as a first step towards a possible
investigation of the `chaotic phenomena' with the method of loop equations.

The paper is organized as follows. In section 2 the basic definitions are
given. Section 3 sets up the loop equation for multi-cut correlators and its
planar solution. In section 4 the iterative solution for two cuts is explained
in detail, and explicit results for genus one and two are obtained. The last
section before concluding is devoted to the double-scaling limit.

\sect{Basic Definitions}

\indent

The complex one-matrix model is defined by the partition function
\be
Z \ [N,\{g_i\}] \ = \ e^{N^2 F} \ =\ \int d\phi^{\dag} d\phi \ \mbox{exp}
(-N \ \mbox{Tr} V(\PP)) \label{Z}
\ee
with
\be
V(\PP) \ = \ \sum_{j=1}^\infty \frac{g_j}{j}(\PP)^j \ ,
\ee
where the integration is over complex $N \times N$ matrices. The generating
functional or one-loop average is given by
\be
W(p) \ = \
\frac{1}{N} \sum_{k=0}^\infty \frac{\langle \mbox{Tr}(\PP)^k \rangle}{p^{2k+1}}
  \ = \ \frac{1}{N} \left\langle \mbox{Tr} \frac{p}{p^2-\PP} \right\rangle \ .
\ee
Introducing the loop insertion operator
\be
\dV(p) \ \equiv \ - \sum_{j=1}^\infty \frac{j}{p^{2j+1}} \frac{d}{dg_j}
\ee
the generating functional can be obtained from the free energy
\be
W(p)\ = \ \dV (p) F + \frac{1}{p} \ . \label{dF1}
\ee
More generally one gets the $n$-loop correlator by iterative application
of the loop insertion operator to $F$,
\be
W(p_1,\ldots,p_n) \ = \ \dV(p_n)\dV(p_{n-1})\cdots\dV(p_1) F \ ,\  n \ge 2 \ ,
\label{dF}
\ee
where
\be
W(p_1,\ldots,p_n) \ \equiv \ \sum_{k_1,\ldots,k_n=1}^\infty
\frac{\langle \mbox{Tr} (\PP)^{k_1} \cdots \mbox{Tr}(\PP)^{k_n}\rangle_{conn}}
{p_1^{2k_1+1} \cdots p_n^{2k_n+1}}  \label{Wdef}
\ee
and $conn$ refers to the connected part.
As the multi-loop correlators and the free energy have the same $1/N$
expansion
\bea
W(p_1,\ldots,p_n) &=& \sum_{g=0}^\infty \frac{1}{N^{2g}} W_g(p_1,\ldots,p_n)
                                                        \label{Wg} \ ,\\
F &=&  \sum_{g=0}^\infty \frac{1}{N^{2g}} F_g \ ,
\eea
relation (\ref{dF}) holds for each genus separately.

\sect{The Loop Equation} \label{loopE}

\indent
In this section the loop equation for the complex matrix model \cite{MA90}
and its planar solution will be given. The explicit formulas are restricted
to the two-cut case for simplicity and for consistency with the
following sections. The multi-cut case can be found in the appendix.
The origin of technical difficulties for more than two cuts will be
also explained in this section.

The form of the loop equation depends explicitly on the number of cuts of
the one-loop correlator only via the contour $\cal C$ of the complex
integral (see appendix \ref{A}). One has
\be
\cI \frac{\om V^{\prime}(\om)}{p^2-\om^2} W(\om) \ = \
   (W(p))^2 + \frac{1}{N^2}\dV(p)W(p) \ \ , \ p\not \in \sigma \ .
   \label{loop}
\ee
Here, $V(\om) = \sum_{j=1}^\infty \frac{g_j}{j}\om^{2j}$, and the support
of the eigenvalue density in the two-cut case is $\sigma= [-x,-y]\cup [y,x]$.
The contour $\cal C$ depicted in Fig.1 encloses $all$ eigenvalues in
such a way that $p$ can also take values on the open real interval
between the cuts of $W(p)$. The generalization to $s$ cuts is obvious.

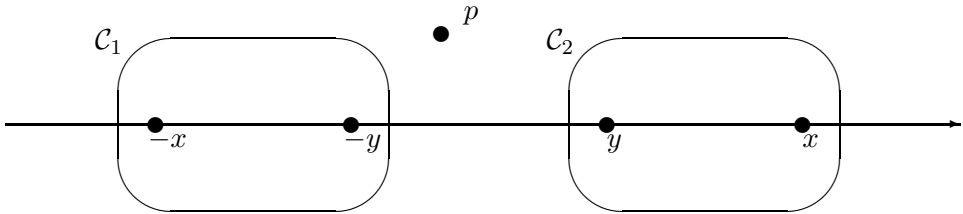
\begin{figure}[h]
\unitlength1cm
\begin{picture}(12.2,5.5)
\thicklines
\multiput(2.1,2.3)(6,0){2}{\line(1,0){2.4}}
\thinlines
\put(0,2.3){\line(1,0){1.9}}
\put(4.7,2.3){\line(1,0){3.2}}
\put(10.7,2.3){\vector(1,0){2}}
\multiput(2,2.3)(2.6,0){2}{\circle*{0.2}}
\multiput(8,2.3)(2.6,0){2}{\circle*{0.2}}
\put(5.8,3.5){\circle*{0.2}}
\put(6.1,3.7){$p$}
\put(1.9,2){$-x$}
\put(4.5,2){$-y$}
\put(8,2){$y$}
\put(10.6,2){$x$}
\multiput(3.3,2.3)(6,0){2}{\oval(3.6,2.3)}
\put(1.2,3.3){${\cal C}_{1}$}
\put(7.2,3.3){${\cal C}_{2}$}
\end{picture}
\caption{the integration contour
         $ {\cal C} = {\cal C}_{1} \cup {\cal C}_{2} $ }
                     \label{fig1}
\end{figure}

Now the iterative solution of (\ref{loop}) works as follows
\cite{ACM92,AKM92,AMB93}.
First the planar solution $W_0(p)$ is determined by taking the limit
$N\rightarrow \infty$, omitting the last term on the r.h.s.. This solution
will then be used as a starting point for the iteration, which calculates
$W_g(p)$ step by step from terms of lower genera. For the case of two cuts,
$W_0(p)$ is given by
\be
W_0(p) \ = \ \frac{1}{2}\cI\frac{pV^{\prime}(\om)}{p^2-\om^2}
             \frac{\pho(\om)}{\pho(p)}
 \label{W0}
\ee
with
\be
\pho(\om) \ \equiv \ \frac{1}{\sqrt{(\om^2-x^2)(\om^2-y^2)}} \ .\label{ph0}
\ee
The extension to $s$ cuts can be found in appendix \ref{B}. Equation (\ref{W0})
differs from the one-cut solution not only by the second square root
but also by a factor of $p$ instead of $\om$ in the numerator.
This stems from the fact that depending on the number of cuts $s$ being even
or odd the complex function $\sqrt{(\om^2-x_1^2)\cdots(\om^2-x_s^2)}$ is to
be defined as an even or odd function of $\om$ respectively (see also
appendix \ref{B}).

{}From the unit normalization of the eigenvalue density it follows that
\be
\lim_{p\to\infty}W(p) \ = \ \frac{1}{p} \ . \label{Winf}
\ee
The leading asymptotic term must be accounted for already by the planar
solution $W_0(p)$ as $1/p$ does not depend on $N$. Imposing this condition
on eq. (\ref{W0}) one finds
\be
\delta_{k,2} \ = \ \frac{1}{2}
 \cI \om^k V^{\prime}(\om) \pho(\om) \ ,\ k=0 \ \mbox{and} \ 2 \ ,
 \label{xy}
\ee
which implicitly determines $x$ and $y$ as functions of the coupling
constants $g_i$.

At this point it should be mentioned that for more than two cuts the condition
(\ref{Winf}) does not supply any more enough equations to determine all
endpoints of the cuts.
In the complex matrix model with an $s$-cut solution there are $s$
such parameters $x_i$ to be determined. In eq. (\ref{xy}) $k$ then runs
over $s,s-2,s-4,\ldots $ down to 0 or 1 depending on whether $s$ is even or
odd. So this yields only $(s+2)/2$ or $(s+1)/2$ equations for $s$
even or odd. The missing equations can be derived from the requirement that
the chemical potentials between the cuts are equal \cite{JU90}, namely
\be
\int_{\bar{\sigma_i}}d\lambda
          \ \rho (\lambda ) \ = \ 0 \ , \ i=1,\ldots,s-1 \ . \label{chem}
\ee
The eigenvalue density $\rho (\lambda )$ is given in the next section and the
$\bar{\sigma_i}$ are the bounded connected components of the real complement
of the support $\sigma$ of $\rho (\lambda )$. Because of symmetry only the
intervals $\bar{\sigma_i}$ on the positive real line need to be considered,
which provides the remaining $(s-2)/2$  or $(s-1)/2$ equations.
They lead to a more complicated dependence of the $x_i$ containing
elliptic integrals, except in the case of $s=1$ or $2$ where they are
trivially fulfilled. To see this the loop insertion operator is applied
to eq. (\ref{chem}), which yields
\be
0 \ = \ M_1^{(i)}\frac{dx_i^2}{dV}(p)
      \int_{\bar{\sigma_j}}d\lambda
      \frac{\sqrt{(\lambda^2-x_1^2)\cdot\ldots\cdot (\lambda^2-x_s^2)}}
{(\lambda^2-x_i^2)}\cdot \left\{
            \begin{array}{rl}
            \lambda & s \ \mbox{even} \\
            1 & s \ \mbox{odd .}
            \end{array}  \right.
\ee
The moments $M_1^{(i)}$ are to be defined in the next section.
Together with $\dV \scr (p)$ of eq. (\ref{xy}) the relations (\ref{chem})
determine the derivatives $\frac{dx_i^2}{dV}\scr (p)$ for $i=1,\ldots,s$,
which are needed in the iterative process. However, their complicated
structure makes it hard to see whether a scheme for calculating higher
genera can still be established.

Coming back to the iterative solution of the loop equation it turns out that
after the insertion of the genus expansion eq. (\ref{Wg}), $W_g(p)$ is
determined by the following equation
\be
(\hat{K}-2W_0(p))W_g(p) \ = \ \sum_{g\prime=1}^{g-1}W_{g\prime}(p)
          W_{g-g\prime}(p)+\dV(p)W_{g-1}(p) \ , \ g\ge 1 \label{gloop} \ .
\ee
Here, the linear operator $\hat{K}$ is defined by
\be
\hat{K} f(p) \ \equiv \ \cI \frac{\om V^{\prime}(\om)}{p^2-\om^2}f(\om) \ .
\ee
$W_g(p)$ is now expressed completely in terms of $W_{g\prime}(p)$ with
$g\prime < g$ on the r.h.s. of eq. (\ref{gloop}). Hence the next step is
the inversion of the operator $(\hat{K}-2W_0(p))$ acting on it. In contrast to
the one-cut case this operation will involve zero modes
contributing to $W_g(p)$ which have to be fixed.

\sect{The iterative solution}

\subsection{Change of variables} \label{Var}

\indent

In analogy to \cite{AMB93} it is convenient to change variables
from the coupling
constants $g_i$ to the moments $M_k$ and $J_k$ in the following way
\bea
M_k &\equiv&  \cI \frac{V^{\prime}(\om) \pho(\om)}{(\om^2-x^2)^k} \ , \
   k\ge1 \ ,\nonumber  \\
J_k &\equiv& \cI \frac{V^{\prime}(\om) \pho(\om)}{(\om^2-y^2)^k} \ , \
   k\ge1 \ . \label{Mom}
\eea
The advantage of these new variables is that, for given genus, $F_g$ and the
multi-loop correlators $W_g(p_1,\ldots,p_n)$ depend only on a finite number
of moments instead of the infinite set of couplings.
Moreover, the $m$th multi-critical point can be characterized by the vanishing
of the first $m-1$ moments $M_k$ or $J_k$. This happens whenever
extra zeros of the eigenvalue density
\be
\rho(\lambda) \ = \ \frac{1}{\pi}\left| {\cal I} \mbox{m}W_0(\lambda) \right|
              \ = \ \frac{1}{4\pi}\left| M(\lambda)\right|
          \sqrt{(x^2-\lambda^2)(\lambda^2-y^2)} \ ,\ \lambda \in \sigma \ .
\ee
occur at either $x^2$ or $y^2$.
The analytic part $M(\lambda)$, given in appendix \ref{B}, must develop these
extra zeros. Using eqs. (\ref{Mfin}) and (\ref{Mom}) it can be seen that the
moments defined above provide an expansion of $M(\lambda)$, i.e.
\be
M(\lambda) \ = \ \sum_{k=1}^\infty \lp M_k \lambda (\lambda^2-x^2)^{k-1}
                                  + J_k \lambda (\lambda^2-y^2)^{k-1} \rp \ .
\ee
Calculating the moments in terms of the coupling constants yields
\bea
M_k &=& g_{k+1} + \lp(k+\frac{1}{2})x^2+\frac{1}{2}y^2\rp g_{k+2}+ \ldots \ ,
                                                      \nonumber \\
J_k &=& g_{k+1} + \lp(k+\frac{1}{2})y^2+\frac{1}{2}x^2\rp g_{k+2}+ \ldots \ .
                                                      \label{Mgk}
\eea

\indent

\subsection{Inversion of $(\hat{K}-2W_0(p))$} \label{sectK}

\indent

In order to proceed in solving equation (\ref{gloop}) it is necessary
to find a set of basis functions for the operator $(\hat{K}-2W_0(p))$ acting
on $W_g(p)$ as follows
\bea
(\hat{K}-2W_0(p))\ \chi^{(n)}(p) &=& \frac{1}{(p^2-x^2)^n} \ , \ n\ge 1 \ ,
                                       \nonumber \\
(\hat{K}-2W_0(p))\ \psi^{(n)}(p) &=& \frac{1}{(p^2-y^2)^n} \ , \ n\ge 1 \ .
                                       \label{Kb}
\eea
As the main result, $W_g(p)$ will then be expressed in terms of these
functions,
\be
W_g(p) \ =  \ \sum_{n=1}^{3g-1} A_g^{(n)}\chi^{(n)}(p) \ +
                                B_g^{(n)}\psi^{(n)}(p) \ , \label{resultW}
\ee
where the coefficients $A_g^{(n)}$ and $B_g^{(n)}$ only depend on the
moments $M_k$ and $J_k$ and on $x^2$ and $y^2$.

Note that eq. (\ref{Kb}) does not yet completely determine $\chi^{(n)}(p)$
and $\psi^{(n)}(p)$ due to a non-trivial kernel of $(\hat{K}-2W_0(p))$.
Because of eq. (\ref{Winf}) only terms asymptotically of order
$p^{-k}, k\ge 2$,
can contribute to $W_g(p)$ for $g\ge1$. In the one-cut case this made the
definition of the basis unique in a simple way.
The argument excluded that the zero mode
$1/\sqrt{p^2-x^2}$ of $(\hat{K}-2W_0(p))_{one-cut}$, which is asymptotically
of order $\frac{1}{p}$, could be added to $W_g(p)$.
However, here this is no longer the case as
\be
\mbox{Ker}(\hat{K}-2W_0(p))\ = \ \mbox{Span}
               \left\{ p\pho(p),\ \frac{1}{p}\pho(p) \right\}
\ee
and
\be
\frac{\pho(p)}{p} \ \sim \ \frac{1}{p^3} \ .
\ee
Hence such a term can be added to $W_g(p)$ in every step of the iteration.
How shall $\chi^{(n)}(p)$ and $\psi^{(n)}(p)$ be fixed such that
$W_g(p)$ is uniquely determined? By definition (see eq. (\ref{dF1}))
$W_g(p)$ can be written as a total derivative
\be
W_g(p) \ = \ \dV(p) \ F_g \ ,\ g\ge1 \ . \label{Wege}
\ee
In order to satisfy this equation, the $p$-dependence of $\chi^{(n)}(p)$ and
$\psi^{(n)}(p)$ must be completely absorbed in terms of the type
$\frac{dx^2}{dV}\scr (p)$, $\frac{dy^2}{dV}\scr (p)$,
$\frac{dM_k}{dV}\scr (p)$ and $\frac{dJ_k}{dV}\scr (p)$.
Consequently the basis functions $\chi^{(n)}(p)$ and
$\psi^{(n)}(p)$ must be linear combinations of these, the
coefficients of course again depending on moments, $x^2$ and $y^2$.
In this manner
the zero mode contributions to the basis, which are indeed necessary,
become uniquely fixed. The final result, proven in appendix \ref{C}, takes
the following form
\bea
\chi^{(n)}(p) &\equiv& \frac{1}{M_1}\lp
   \frac{1}{x^2}\Big(\phi_x^{(n)}(p)-\sum_{k=1}^{n-1}\chi^{(k)}(p)M_{n-k}\Big)
   - \sum_{k=1}^{n-1}\chi^{(k)}(p)M_{n-k+1} \rp \ , \nonumber \\
\psi^{(n)}(p) &\equiv& \frac{1}{J_1} \lp
   \frac{1}{y^2}\Big(\phi_y^{(n)}(p)-\sum_{k=1}^{n-1}\psi^{(k)}(p)J_{n-k}\Big)
        - \sum_{k=1}^{n-1}\psi^{(k)}(p)J_{n-k+1} \rp , \label{basis}
\eea
where
\bea
\phi_x^{(n)}(p) &\equiv& \frac{p \pho(p)}{(p^2-x^2)^n} \ ,\ n\ge1  \ ,
                                                      \nonumber \\
\phi_y^{(n)}(p) &\equiv& \frac{p \pho(p)}{(p^2-y^2)^n}  \ ,\ n\ge1 \ .
\eea

\indent

\subsection{The iterative procedure determining $W_g(p)$ }

\indent

In order to solve eq. (\ref{gloop}), one first has to compute
$\dV\scr (p)$$W_0(p)=W_0(p,p)$. To do so it is convenient to rewrite
the loop insertion operator
\be
\dV (p) \ = \ \parV (p) +\frac{dx^2}{dV}(p)\frac{\partial}{\partial x^2}
                        +\frac{dy^2}{dV}(p)\frac{\partial}{\partial y^2}
                        \label{newdV} \ ,
\ee
where
\be
\parV (p) \ \equiv \ -\sum_{j=1}^{\infty} \frac{j}{p^{2j+1}}
                                          \frac{\partial}{\partial g_j} \ .
\ee
For evaluating $\frac{dx^2}{dV}\scr (p)$ and $\frac{dx^2}{dV}\scr (p)$,
eq. (\ref{newdV}) is applied to eq. (\ref{xy}) which yields
\be
\frac{dx^2}{dV}(p) \ = \ \frac{1}{M_1} \phi_x^{(1)}(p) \ , \ \
\frac{dy^2}{dV}(p) \ = \ \frac{1}{J_1}\ \phi_y^{(1)}(p) \ .
\footnote{This simple relationship with the basis functions is
spoiled for more than two cuts.}
                                                            \label{dxy}
\ee
Now with eq. (\ref{newdV}) $\dV\scr (p)$$W_0(p)$ can be calculated. Using
the relation
\be
\parV (p) \Vp (\om) \ = \ -\frac{2\om p}{(p^2-\om^2)^2} \ ,
\ee
the definitions (\ref{Mom}) and deforming the contour of the remaining integral
to infinity, one gets
\be
W_0(p,p) \ = \ \frac{p^2d^2}{16(p^2-x^2)^2(p^2-y^2)^2} \ ,
\ee
where
\be
d \ \equiv \ x^2-y^2.
\ee
Before one can make use of the basis functions eq. (\ref{basis}) for
$(\hat{K}-2W_0(p))$ one has to
decompose the r.h.s. of eq. (\ref{gloop}) into fractions of the form
$(p^2-x^2)^{-n}$ and $(p^2-y^2)^{-n}$ . Doing so the coefficients $A_g^{(n)}$
and $B_g^{(n)}$ in (\ref{resultW}) can now be identified for genus 1.
\bea
A_1^{(1)} &=& \frac{1}{16}-\frac{1}{8}\frac{x^2}{d} \ , \ \
              A_2^{(1)} \ = \ \frac{x^2}{16} \nonumber \\
B_1^{(1)} &=& \frac{1}{16}+\frac{1}{8}\frac{y^2}{d} \ , \ \
              B_2^{(1)} \ = \ \frac{y^2}{16} \label{gen1}
\eea
It is clear how to carry on the iteration process. However, in order to
calculate $\dV\scr (p)$$W_g(p)$ it is convenient to rewrite the loop
insertion operator once again,
\bea
\dV (p) &=& \sum_{n=1}^{\infty} \frac{dM_n}{dV}(p)\frac{\partial}{\partial M_n}
             +\sum_{n=1}^{\infty} \frac{dJ_n}{dV}(p)
                                  \frac{\partial}{\partial J_n}
             +\frac{dx^2}{dV}(p)\frac{\partial}{\partial x^2}
             +\frac{dy^2}{dV}(p)\frac{\partial}{\partial y^2} \ ,
                                                 \nonumber \\
                                                 \label{dVMJ} \\
\frac{dM_n}{dV}(p) &=&
       -(n+\frac{1}{2})\phi_x^{(n+1)}(p)
           -\frac{1}{2}\sum_{k=1}^n \frac{(-1)^{k+n}}{d^{n-k+1}}
              \lp \phi_x^{(k)}(p)-M_k \frac{dy^2}{dV}(p) \rp \nonumber \\
          && +(n+\frac{1}{2}) M_{n+1}\frac{dx^2}{dV}(p) \ .
     \label{dM}
\eea
With the definition (\ref{Mom}) of the moments and the form
of $\dV\scr (p)$ in eq. (\ref{newdV}) the relation is easily verified.
The corresponding result for $\frac{dJ_n}{dV}\scr (p)$ can be obtained
by interchanging $x^2 \leftrightarrow y^2$ and $M_k \leftrightarrow J_k$.
Looking back at the original loop equation (\ref{loop}) it is obvious
that the free energy and the multi-loop correlators for all genera
should be invariant under such an interchangement .

Using the loop insertion operator in the form of eq. (\ref{dVMJ}) a lengthy
calculation yields the following result for $g=2$:
\bea
A_2^{(1)} &=& -\frac{5}{32}   \frac{y^2M_3}{d^3(M_1)^3}
           \  -\frac{5}{32}   \frac{x^2J_3}{d^3(J_1)^3}
              -\frac{1}{64}   \frac{x^2M_2J_2}{d^3(M_1)^2(J_1)^2}
              +\frac{1}{128}  \frac{M_2J_2}{d^2(M_1)^2(J_1)^2} \nonumber \\
           && +\frac{49}{256} \frac{y^2(M_2)^2}{d^3(M_1)^4}
              +\frac{49}{256} \frac{x^2(J_2)^2}{d^3(J_1)^4}
              +\frac{11}{128} \frac{x^2M_2}{d^4(M_1)^2J_1}
              -\frac{11}{128} \frac{y^2J_2}{d^4M_1(J_1)^2}     \nonumber \\
           && +\frac{67}{128} \frac{y^2M_2}{d^4(M_1)^3}
              -\frac{67}{128} \frac{x^2J_2}{d^4(J_1)^3}
              -\frac{3}{128}  \frac{M_2}{d^3(M_1)^2J_1}
              -\frac{3}{128}  \frac{J_2}{d^3M_1(J_1)^2}        \nonumber \\
           && +\frac{201}{256}\frac{y^2}{d^5(M_1)^2}
              +\frac{201}{256}\frac{x^2}{d^5(J_1)^2}
              +\frac{57}{64}  \frac{x^2}{d^5M_1J_1}
              -\frac{57}{128} \frac{1}{d^4M_1J_1}   \ ,        \nonumber \\
A_2^{(2)} &=& +\frac{5}{32}   \frac{y^2M_3}{d^2(M_1)^3}
              +\frac{1}{128}  \frac{x^2M_2J_2}{d^2(M_1)^2(J_1)^2}
              -\frac{49}{256} \frac{y^2(M_2)^2}{d^2(M_1)^4} \nonumber \\
           && -\frac{3}{128}  \frac{x^2M_2}{d^3(M_1)^2J_1}
              +\frac{1}{16}   \frac{x^2J_2}{d^3M_1(J_1)^2}
              -\frac{3}{128}  \frac{J_2}{d^2M_1(J_1)^2}
              -\frac{67}{128} \frac{y^2M_2}{d^3(M_1)^3}     \nonumber \\
           && -\frac{57}{128} \frac{x^2}{d^4M_1J_1}
          \ \ -\frac{201}{256}\frac{y^2}{d^4(M_1)^2}
              +\frac{15}{128} \frac{1}{d^3M_1J_1}      \ ,   \nonumber \\
A_2^{(3)} &=& -\frac{5}{32}  \frac{x^2M_3}{d(M_1)^3}
              +\frac{49}{256}\frac{x^2(M_2)^2}{d(M_1)^4}
              -\frac{5}{128} \frac{x^2J_2}{d^2M_1(J_1)^2}
              +\frac{67}{128}\frac{x^2M_2}{d^2(M_1)^3}      \nonumber \\
           && -\frac{49}{128} \frac{M_2}{d(M_1)^3}
              +\frac{15}{128} \frac{x^2}{d^3M_1J_1}
              +\frac{201}{256}\frac{x^2}{d^3(M_1)^2}
            \ -\frac{189}{256}\frac{1}{d^2(M_1)^2}   \ ,    \nonumber \\
A_2^{(4)} &=& -\frac{49}{128} \frac{x^2M_2}{d(M_1)^3}
              -\frac{189}{256}\frac{x^2}{d^2(M_1)^2}
              +\frac{105}{256}\frac{1}{d(M_1)^2}      \ ,   \nonumber \\
A_2^{(5)} &=& \frac{105}{256} \frac{x^2}{d(M_1)^2}    \ ,   \nonumber \\
                                                            \nonumber \\
B_2^{(i)} &=& A_2^{(i)}(M_k\leftrightarrow J_k,x^2 \leftrightarrow y^2)
          \ , \ i=1,\ldots,5 \ . \label{gen2}
\eea
$W_2(p)$ is then obtained by inserting eqs.
(\ref{basis}) and (\ref{gen2}) into eq. (\ref{resultW}).

\indent

\subsection{The iterative procedure for $F_g$ }

\indent

As it was already pointed out in section \ref{sectK} and proven in
appendix \ref{C} the basis functions $\chi^{(n)}(p)$ and
$\psi^{(n)}(p)$ can be
written as linear combinations of derivatives with respect to $V(p)$.
This requirement was imposed in order to fix the basis in a unique
way. For genus 1 the explicit result reads
\bea
\chi^{(1)}(p) &=& \frac{1}{x^2}\frac{dx^2}{dV}(p) \ ,\nonumber \\
\chi^{(2)}(p) &=& -\frac{2}{3}\frac{1}{x^2M_1}\frac{dM_1}{dV}(p)
                  -\frac{1}{x^4}\frac{dx^2}{dV}(p)
                  -\frac{1}{3}\frac{1}{x^2}\dV (p) \ln(d) \ ,\nonumber \\
\psi^{(1)}(p) &=& \frac{1}{y^2}\frac{dy^2}{dV}(p) \ ,\nonumber \\
\psi^{(2)}(p) &=& -\frac{2}{3}\frac{1}{y^2J_1}\frac{dJ_1}{dV}(p)
                  -\frac{1}{y^4}\frac{dy^2}{dV}(p)
                  -\frac{1}{3}\frac{1}{y^2}\dV (p) \ln(d) \ ,\label{dbasis}
\eea
which can easily be verified from the definitions. The relation (\ref{Wege})
then allows to calculate $F_g$ for any
given $W_g(p)$. So from eq. (\ref{dbasis}) in combination with eqs.
(\ref{gen1}) and (\ref{resultW}) $F_1$ can now be read off up to a constant
\be
F_1 \ = \ -\frac{1}{24}\ln(M_1) - \frac{1}{24}\ln(J_1) -\frac{1}{6}\ln(d)\ .
\ee
Continuing in the same manner and rewriting the basis as being sketched
in appendix \ref{C} the genus two result is obtained using eq. (\ref{gen2})
after some tedious work,
\bea
F_2
&=& -\frac{35}{384}\frac{M_4}{d(M_1)^3} +\frac{35}{384}\frac{J_4}{d(J_1)^3}
    +\frac{43}{192}\frac{M_3}{d^2(M_1)^3}+\frac{43}{192}\frac{J_3}{d^2(J_1)^3}
                                                    \nonumber \\
&&  +\frac{29}{128}\frac{M_3M_2}{d(M_1)^4}
    -\frac{29}{128}\frac{J_3J_2}{d(J_1)^4}
    +\frac{1}{64}\frac{M_2J_2}{d^2(M_1)^2(J_1)^2}   \nonumber \\
&&  -\frac{21}{160}\frac{(M_2)^3}{d(M_1)^5}
    +\frac{21}{160}\frac{(J_2)^3}{d(J_1)^5}
    -\frac{11}{40} \frac{(M_2)^2}{d^2(M_1)^4}
    -\frac{11}{40} \frac{(J_2)^2}{d^2(J_1)^4}       \nonumber \\
&&  -\frac{181}{480}\frac{M_2}{d^3(M_1)^3}
    +\frac{181}{480}\frac{J_2}{d^3(J_1)^3}
    -\frac{3}{64}\frac{M_2}{d^3(M_1)^2J_1}
    +\frac{3}{64}\frac{J_2}{d^3 M_1(J_1)^2}         \nonumber \\
&&  -\frac{181}{480}\frac{1}{d^4(M_1)^2}
    -\frac{181}{480}\frac{1}{d^4(J_1)^2}
    -\frac{5}{16}\frac{1}{d^4M_1J_1} \ .
\eea
{}From eq. (\ref{resultW}) and the procedure described just above it should
have
become obvious that $F_g$ depends on at most $2(3g-2)$ moments.

Another remarkable fact is that the calculated $F_1$ and $F_2$ exactly
coincide with those of the one-cut hermitian matrix model described in
\cite{AMB93} when the identification of the moments $M_k$ and $J_k$ and
the difference $d$ is made (using the same notation). This
coincidence away from the double-scaling limit cannot be merely pure
coincidence. However, since the loop insertion operators are distinct the loop
correlators will differ as it can already be seen for $W_1(p)$ and $W_2(p)$.
Relating the two models it has been mentioned in \cite{DJM92}
that the complex matrix model corresponds to a hermitian matrix model
with a general potential where the eigenvalues are restricted to be positive
\footnote{This can be seen after diagonalisation when the eigenvalues of
$\PP$ are considered to be $\lambda \ge 0$ instead of $\lambda^2$ with
$\lambda$ real here.}.
It seems that the repulsion at the origin ($\lambda=0$) is reflected only
in the multi-loop correlators but not seen in the free energy.

Finally a comparison to the one-cut solution of the complex matrix model
presented in \cite{AKM92} can be made by letting $y\rightarrow 0$.
The moments defined here and there then translate into each other.
\bea
M_k(g_{k+1}\rightarrow g_k) &\rightarrow & {\cal I}_{k} \nonumber \\
J_k&\rightarrow & {\cal M}_{k}
\eea
As one might have expected, the results for the free energy and the multi-loop
correlators do not match except for the universal 2-loop correlator
$W_0(p,p)$. For a given set of couplings the phase boundary between the
one- and two-cut phase can be formulated.
The free energy or the correlators can be compared on the boundary,
inspecting the order of the phase transition and critical exponents.

\indent

\subsection{The general structure of $F_g$ and $W_g(p)$}

\indent

The main result for $F_g$ from the iterative solution of the loop equation
can be written in the following way
\be
F_g \ = \ \sum_{\alpha_i > 1,\beta_j > 1}
      \langle \alpha_1\ldots \alpha_k;\beta_1\ldots \beta_l |
              \alpha,\beta,\gamma \rangle_g \
      \frac{M_{\alpha_1}\ldots M_{\alpha_k}J_{\beta_1}\ldots J_{\beta_l}}
{d^{\gamma}(M_1)^{\alpha}(J_1)^{\beta}} ,\ g\ge 2 \ . \label{Fstr}
\ee
Here the brackets denote rational numbers and $\alpha, \beta$ and $\gamma$ are
non-negative integers. The summation-indices $\alpha_i$ and $\beta_j$ take
values in the interval $[2,3g-2]$. For every genus $g, \ F_g$ contains only
finitely many terms with a finite number of moments. In particular $F_g$
depends on at most $2(3g-2)$ different moments. This structure can either be
proven along the same lines like in \cite{AKM92} or becomes clear when
performing the first steps of the iteration. In perfect analogy to the
one-cut case of the hermitian matrix model \cite{AMB93} several relations
between the indices and powers in eq. (\ref{Fstr}) can be derived.

First of all because of the symmetry
$x^2\leftrightarrow y^2, M_k\leftrightarrow J_k$ the following holds
\be
\langle \alpha_1\ldots \alpha_k;\beta_1\ldots \beta_l |
              \alpha,\beta,\gamma \rangle_g =
 (-1)^{\gamma}\langle \beta_1\ldots \beta_l;\alpha_1\ldots \alpha_k |
              \beta,\alpha,\gamma \rangle_g \ .
\ee
Defining
\be
N_M \ = \ k-\alpha \ ,\ \ N_J \ = \ l-\beta\ , \label{NM}
\ee
it is true that
\be
N_M \ \le \ 0\ ,\ \ N_J \ \le \ 0 \ . \label{NMO}
\ee
The invariance of the partition function $Z=$exp$(\sum_g N^{2-2g}F_g)$
under the rescaling $N\rightarrow kN$ and
$\rho(\lambda)\rightarrow \frac{1}{k}\rho(\lambda)$ for each genus yields
\be
N_M+N_J \ = \ 2-2g \ . \label{NMF}
\ee
The rescaling $N\rightarrow k^2 N$ and $g_j\rightarrow k^{j-2}g_j$ implies
\bea
M_j &\rightarrow & k^{j-1}M_j \ ,\ \ J_j \ \rightarrow \ k^{j-1}J_j
                                          \ , \nonumber \\
x^2 &\rightarrow & k^{-1} x^2 \ \ \ ,\ \ y^2 \ \rightarrow \ k^{-1}y^2
\eea
because of eq. (\ref{Mgk}). For $F_g$ this reads
\be
\sum_{i=1}^k (\alpha_i -1)+\sum_{j=1}^l (\beta_j -1)+\gamma =4g-4 \ .
\ee
In the double-scaling limit in the next chapter further relations of this
type will be derived allowing to decide which terms in eq. (\ref{Fstr}) will
survive.

Turning to $W_g(p)$ the coefficients $A_g^{(n)}$ and $B_g^{(n)}$ have a
similar structure like in eq. (\ref{Fstr}) since $W_g(p)$ follows from $F_g$
by applying the loop insertion operator. In particular,
\be
A_g^{(n)} \ =  \sum_{\alpha_i > 1,\beta_j > 1}
      \langle \alpha_1\ldots \alpha_k;\beta_1\ldots \beta_l |
              \alpha,\beta,\gamma \rangle_g^{(n)} \
 \frac{M_{\alpha_1}\ldots M_{\alpha_k}J_{\beta_1}\ldots J_{\beta_l}}
 {(M_1)^{\alpha}(J_1)^{\beta}} f(x^2,y^2) \label{Astr}
\ee
for $g\ge1$
and analogously for $B_g^{(n)}$. The only difference is that the
indices $\alpha_i$ and $\beta_j$ lie in the interval $[2,3g-n]$.
Because of the same genus expansion eq. (\ref{NM}) is also valid here.
So with the definition (\ref{basis}) of the basis functions $W_g(p)$ depends
on at most $2(3g-1)$ moments.

\sect{The double-scaling limit} \label{dsl}

\indent

In the conventional double-scaling limit all matrix models belonging to the
same universality class should be equivalent. Consequently all differences
originating from the multi-cut structure should vanish in this
limit. Having explicit results at hand for the one-cut hermitian and
the one- and two-cut complex matrix model (\cite{AKM92},\cite{AMB93}),
this can be checked as an example.

In the following the double-scaling limit will be performed for $x^2$
only; the one for $y^2$ is then easily obtained. For the $m$th
multi-critical model the couplings are adjusted such that
\bea
x^2 &=& x_c^2 - a \Lambda^{\frac{1}{m}} \ , \nonumber \\
p^2 &=& x_c^2 + a \pi \label{xsc}
\eea
and $y^2$ does not scale. Here $a$ is the scaling parameter, which becomes
zero at the critical point. As it had been already mentioned in section
\ref{Var} the eigenvalue density then develops $m-1$ extra zeros at $x^2$
and hence
\be
M_k \ \sim a^{m-k} \ ,\ k=1,\ldots ,m-1 \ ,\label{Msc}
\ee
whereas the $J_k$'s do not scale. The resulting contribution to the free
energy has the well known behavior
\be
F_g \ \sim \ a^{(2-2g)(m+\frac{1}{2})} \ , \ g\ge 1.
\ee
Making use of eq. (\ref{Fstr}) for $F_g$ leads to
\be
\sum_{i=1}^k(m-\alpha_i)- \alpha (m-1)\ge m(2-2g)-g+1 \ .
\ee
Since in the scaling limit the free energy should look the same for all
multi-critical models it follows that
\bea
N_M &\ge & 2-2g \ ,\nonumber \\
       \sum_{i=1}^k(\alpha_i-1) &\le & 3g-3 \ .
\eea
{}From eqs. (\ref{NMF}) and (\ref{NMO}) the equality sign holds in the first
equation. Only terms for which this is true also in the second line will
contribute in the double-scaling limit. For example the genus two
contribution to $F$ will then be
\be
F_2^{(d.s.l.)} \ = \ -\frac{35}{384} \frac{M_4}{d(M_1)^3}
                     +\frac{29}{128} \frac{M_3M_2}{d(M_1)^4}
                     -\frac{21}{160} \frac{(M_2)^3}{d(M_1)^5} \ .
\ee

Switching to the one-loop correlator $W_g(p)$ the behavior of the basis
functions in eq. (\ref{basis}) also has to be analyzed. Equations (\ref{xsc})
and (\ref{Msc}) lead to
\bea
\chi^{(n)}(\pi,\Lambda) &\sim & a^{-m-n+\frac{1}{2}} \ ,\nonumber \\
\psi^{(n)}(\pi,\Lambda) &\sim & a^{-\frac{1}{2}} \ .
\eea
The known genus $g$ contribution
\be
W_g(\pi,\Lambda) \ \sim \ a^{(1-2g)(m+\frac{1}{2})-1}
\ee
together with eq. (\ref{Astr}) requires for the $A_g^{(n)}$ terms that
\be
\sum_{i=1}^k(m-\alpha_i)-\alpha(m-1)-m-n+ \frac{1}{2}
                                       \ge m(1-2g)-g- \frac{1}{2}.
\ee
The same argument as above then yields
\bea
N_M &\ge & 2-2g \ , \nonumber \\
         \sum_{i=1}^k(\alpha_i-1) &\le & 3g-n-1 \ ,
\eea
where again only terms obeying equality appear in the scaling limit.
A similar computation for $B_g^{(n)}$ reveals $N_M\ge 1-2g$, which
clearly cannot be fulfilled as an equation together with
eq. (\ref{NMF}) because $N_J\le 0$.
So all the $B_g^{(n)}$ terms will disappear in the double-scaling limit.
Finally the non-vanishing coefficients of $W_2(p)$ are given as an example.
\bea
A_2^{(3)} &=& \frac{49}{256} \frac{x^2_c(M_2)^2}{d_c(M_1)^4}
              -\frac{5}{32}  \frac{x^2_cM_3}{d_c(M_1)^3} \nonumber \\
A_2^{(4)} &=& -\frac{49}{128}\frac{x^2_cM_2}{d_c(M_1)^3} \nonumber \\
A_2^{(5)} &=& \frac{105}{256}\frac{x^2_c}{d_c(M_1)^2}
\eea
In $W_2(p)$ the $x^2_c$-dependence cancels out because in the basis the
second sum is suppressed in the scaling limit.
This reproduces exactly the result for the one-cut hermitian matrix model
in \cite{AMB93}, where the equivalence to the one-cut complex matrix model
had already been proven.

\sect{Conclusions}

\indent

It has been shown how the powerful method of iteratively solving the
loop equation by Ambj{\o}rn $et$ $al$. \cite{ACM92,AKM92,AMB93}
generalizes to the complex matrix
model with more than one cut present. The loop equation  for an arbitrary
number of cuts was derived and solved for the one-loop correlator
in the planar limit. In principle
the procedure to find the genus $g$ contribution is clear also
for the multi-cut type. Nevertheless, for more than two cuts a new kind
of equation determining the edges of the cuts enters and renders the
computation technically much more involved.

The iterative scheme was then explicitly presented for the two-cut model,
and results for genus one and two were obtained away from the double-scaling
limit.
Relations to the one-cut solution of the hermitian and complex matrix
model were discussed, in particular in the case of the double-scaling
limit and when the two cuts merge.

In order to attack the problem of instabilities in multi-cut solutions
termed `chaos in matrix models' a more complicated cut structure has to
be examined, e.g. the hermitian model with two cuts for an arbitrary potential
or simply just with three or more cuts. Up to now the instabilities have only
been found within the picture of orthogonal polynomials. The hope is that
they can be found also in the framework of loop
equations and that a deeper understanding especially concerning
correlation functions can be obtained in this way. These open problems
are left subject to further investigations.

\indent

\begin{flushleft}
\underline{Acknowledgements}: I would like to thank P. Adamietz,
J. Ambj{\o}rn, O. Lech-\\
tenfeld and J. Plefka for helpful discussions.
\end{flushleft}

\begin{appendix}

\sect{Derivation of the loop equation for the $s$-cut solution} \label{A}

\indent

The matrix integral eq. (\ref{Z}) is invariant under the following
transformation
\be
\phi \rightarrow \phi\lp1+\epsilon\frac{p}{p^2-\PP}\rp \ ,\
\phi^{\dagger} \rightarrow \lp 1+\epsilon\frac{p}{p^2-\PP}\rp\phi^{\dagger} \ .
\ee
The functional determinant and the change of the action is then given by
\bea
d\phi d\phi^{\dagger} &\rightarrow&
d\phi d\phi^{\dagger}\lp1+2\epsilon p (\mbox{Tr}\frac{p}{p^2-\PP})^2\rp \ ,\\
V(\PP) &\rightarrow& V(\PP) + 2\epsilon \frac{p\PP}{p^2-\PP}V^{\prime}(\PP)
\ ,\
\eea
where
\be
V^{\prime}(\PP)\ \equiv \ \sum_{n=1}^{\infty} g_n (\PP)^{n-1} \ .
\ee
The invariance of $Z$ then reads
\be
\left\langle (\mbox{Tr}\frac{p}{p^2-\PP})^2\right\rangle
 -N\left\langle \mbox{Tr}\frac{\PP}{p^2-\PP}
   V^{\prime} (\PP)\right\rangle  \ = \ 0 \ ,
\ee
which is already almost the loop equation using eqs. (\ref{dF}) and
(\ref{Wdef})
\be
\frac{1}{N}\left\langle \mbox{Tr}\frac{\PP}{p^2-\PP} V^{\prime}
                                             (\PP)\right\rangle  \ = \
    (W(p))^2 +\frac{1}{N^2} \dV(p)W(p) \ . \label{A1}
\ee
Now the density of the eigenvalues can be formally introduced as
\be
\rho_N(\lambda) \equiv \frac{1}{N} \left\langle \sum_{i=1}^N
                               \delta(\lambda-\lambda_i)\right\rangle \ .
\ee
With $\sigma=\cup_{i=1}^s \sigma_i$ being the support of our $s$-cut solution
the explicit dependence on the number of cuts $s$ enters the l.h.s. of
eq. (\ref{A1}).
\bea
\frac{1}{N} \left\langle \mbox{Tr}\frac{\PP}{p^2-\PP} V^{\prime}
                                                        (\PP)\right\rangle
  &=&  \sum_{i=1}^s \int_{\sigma_i}d\lambda \rho_N(\lambda)
                       \frac{\lambda^2V^{\prime}(\lambda^2)}{p^2-\lambda^2}
                                                         \nonumber   \\
  &=& \sum_{i=1}^s \int_{\sigma_i}d\lambda \rho_N(\lambda)
                   \cI\frac{2\om}{\om^2-\lambda^2}
                   \frac{\om^2 V^{\prime}(\om^2)}{p^2-\om^2}
                                                         \nonumber \\
  &=&\cI W(\om)\frac{2\om^2 V^{\prime}(\om^2)}{p^2-\om^2} \ .
\eea
Here $\cal C$ encloses all cuts without containing $\pm p$, which generalizes
the two-cut case depicted in Figure \ref{fig1} of section \ref{loopE}. With
the following change of notation eq. (\ref{loop}) is finally obtained.
\be
2\om V^{\prime}(\om^2)=2\sum_{n=1}^{\infty}g_n \om^{2n-1} \equiv
V^{\prime}(\om)  \ .
\ee

\sect{The planar solution of the loop equation} \label{B}

\indent

In the limit of $N \rightarrow \infty$ the loop equation (\ref{loop}) becomes
\be
\cI \frac{\om V^{\prime}(\om)}{p^2-\om^2} W_0(\om) \ = \ (W_0(p))^2 \ .
\ee
Deforming the contour $\cal C$ to infinity and using the fact that $W(p)$ and
$V^{\prime}(p)$ are odd functions by definition one gets the following
contributions from the poles at $\pm p$ and $\infty$
\be
(W_0(p))^2 \ = \ \frac{1}{2}V^{\prime}(p) W_0(p)
 +\oint_{\cal C_{\infty}}\frac{d\om}{4\pi i}
      \frac{\om V^{\prime}(\om)}{p^2-\om^2}W_0(\om) \ .
\ee
The solution of this quadratic equation for $W_0(p)$ of course reads
\be
W_0(p) \ = \ \frac{1}{4}\Vp(p) \pm \sqrt{\frac{1}{16}(\Vp(p))^2+Q(p)} \ ,
\ee
with
\be
Q(p) \ = \ \oint_{\cal C_{\infty}}\frac{d\om}{4\pi i}
      \frac{\om V^{\prime}(\om)}{p^2-\om^2}W_0(\om)
\ee
to be calculated for any given potential with finitely many couplings.
Making an ansatz for a solution with $s$ cuts $W_0(p)$ looks the following
\be
W_0(p) \ = \ \frac{1}{4}\lp \Vp(p)-M(p)\sqrt{(p^2-x_1^2)\ldots(p^2-x_s^2)}\rp
\ , \label{B1}
\ee
where $M(p)$ is an analytic function. So
\be
M(p) \ = \  \pho(p) (\Vp(p)-4W_0(p))  \label{M} \ ,
\ee
remembering
\be
\pho(p) =  \frac{1}{\sqrt{(p^2-x_1^2)\ldots(p^2-x_s^2)}}
    \ \equiv \ p^{-s}\lp 1-\frac{x_1^2}{p^2}\rp ^{-\frac{1}{2}}\ldots
           \lp 1-\frac{x_s^2}{p^2}\rp ^{-\frac{1}{2}}.
\ee
Now for $s$ even (odd) $\pho(p)$ is defined as an even (odd) complex function
of $p$ and consequently
\bea
M(p) &=& \frac{1}{2}\lp M(p) +(-1)^{s+1}M(-p)\rp \nonumber \\
     &=& \frac{1}{2}\oint_{\cal C_{\infty}}\frac{d\om}{2\pi i}
          M(\om) \lp\frac{1}{\om-p}+\frac{(-1)^{s+1}}{\om+p}\rp \ . \label{Mi}
\eea
Reinserting eq. (\ref{M}) into the integral in eq. (\ref{Mi}) the term
proportional to $W_0(p)$ drops out because of its asymptotic eq. (\ref{Winf})
and hence
\be
M(p) \ = \ \oint_{\cal C_{\infty}} \frac{d\om}{4\pi i}
\frac{2\Vp(\om)}{\om^2-p^2}\pho(\om)\cdot \left\{
                \begin{array}{rl}
                 p & s \ \mbox{even}\\ \om & s \ \mbox{odd .}
                \end{array} \right.   \label{Mfin}
\ee
Plugging this into eq. (\ref{B1}) again after a similar calculation $W_0(p)$
can then be expressed as
\be
W_0(p) \ = \ \frac{1}{2}\cI \frac{\Vp(\om)}{p^2-\om^2}
                \frac{\pho(\om)}{\pho(p)} \cdot \left\{
                \begin{array}{rl}
                 p & s \ \mbox{even}\\ \om & s \ \mbox{odd .}
                \end{array} \right.
\ee

\sect{Determination of the basis} \label{C}

\indent

As being described in section \ref{sectK}
our aim is to find a basis which can be
expressed completely  in terms of moments, $x^2$ and $y^2$ as well as total
derivatives $\dV\scr (p)$ of them. The results for the latter were already
given in eqs. (\ref{dxy}) and (\ref{dM}).
It will be shown by induction that the basis defined in eq. (\ref{basis})
can be expressed in the conjectured way. The starting point was made in
eq. (\ref{dbasis}).
Now assume that this holds for all $\chi^{(k)}(p) \ , \ k=1,\ldots,n-1$.
Rearranging eq. (\ref{basis}) shows that the same is true then for the
$\phi_x^{(k)}(p)$
\be
\phi_x^{(k)}(p) \ = \ x^2\sum_{l=1}^k \chi^{(l)}(p)M_{k-l+1} +
               \sum_{l=1}^{k-1} \chi^{(l)}(p)M_{k-l} \ , \ k=1,\ldots,n-1 .
\ee
It follows with eq. (\ref{dM}) that the remaining term in $\chi^{(n)}(p)$ can
also be rewritten in the desired way.
\bea
\phi_x^{(n)}(p)
&=& \frac{1}{n-\frac{1}{2}} \lp -\frac{dM_{n-1}}{dV}(p)
   -\frac{1}{2}\sum_{k=1}^{n-1}\frac{(-1)^{n+k-1}}{d^{n-k}}
   \Big( \phi_x^{(k)}(p)-M_k\frac{dy^2}{dV}(p) \Big) \rp \nonumber \\
&&+ \frac{dx^2}{dV}(p)M_n
\eea
The proof for $\psi^{(n)}(p)$ is going exactly along the same lines.
The fixing of the basis is unique because the zero mode
alone cannot be written as a derivative with respect to $V(p)$.

What remains to show is that $\chi^{(n)}(p)$ and $\psi^{(n)}(p)$ really form
a basis like in eq. (\ref{Kb}). First define in analogy with the one-cut case
\be
\tilde{\chi}^{(n)}(p) \ \equiv \ \frac{1}{M_1}
 \lp\frac{1}{p^2}\phi_x^{(n)}(p)-\sum_{k=1}^{n-1}\tilde{\chi}^{(k)}(p)
                                                         M_{n-k+1} \rp \ ,
\ee
which does not contain the zero modes yet. It is easily proven by induction
that
\be
(\hat{K}-2W_0(p))\tilde{\chi}^{(n)}(p) \ = \ \frac{1}{(p^2-x^2)^n} \ ,
                                               \ n\ge1\ ,   \label{C1}
\ee
holds. Now the zero modes are added to $\tilde{\chi}^{(n)}(p)$,
\be
\chi^{(n)}(p) \ \equiv \ \frac{1}{M_1}
 \lp\frac{1}{p^2}\phi_x^{(n)}(p) +\frac{(-1)^{n+1}}{x^{2n}}\frac{\pho(p)}{p}
-\sum_{k=1}^{n-1} \chi^{(k)}(p)M_{n-k+1} \rp  \ , \label{C5}
\ee
so eq. (\ref{C1}) without tilde is still valid.
Finally the equivalence of eq. (\ref{C5}) to the form in eq. (\ref{basis})
is again shown by induction. Proceeding in the same way for $\psi^{(n)}(p)$
completes the proof of this section.

\newpage

\end{appendix}

\end{document}